\documentclass[aps,prl,twocolumn,showpacs,groupedaddress,superscriptaddress,footinbib]{revtex4-2}

\usepackage{mathrsfs}
\usepackage{natbib,hyperref}
\setcitestyle{square,sort&compress,comma,numbers}
\hypersetup{colorlinks=true, citecolor=blue, urlcolor=blue, linkcolor=blue}
\usepackage[english]{babel}
\usepackage[utf8]{inputenc}
\usepackage{graphicx}
\usepackage[caption=false]{subfig}
\usepackage{amsmath}
\usepackage{amsfonts}
\usepackage{amssymb}
\usepackage{soul}
\usepackage{cancel}
\setulcolor{red}
\usepackage{easyReview}
\usepackage{xcolor}

\usepackage{pdfpages} 
\usepackage{pgffor} 

\makeatletter
\AtBeginDocument{\let\LS@rot\@undefined}
\makeatother

\def\supplementfilename{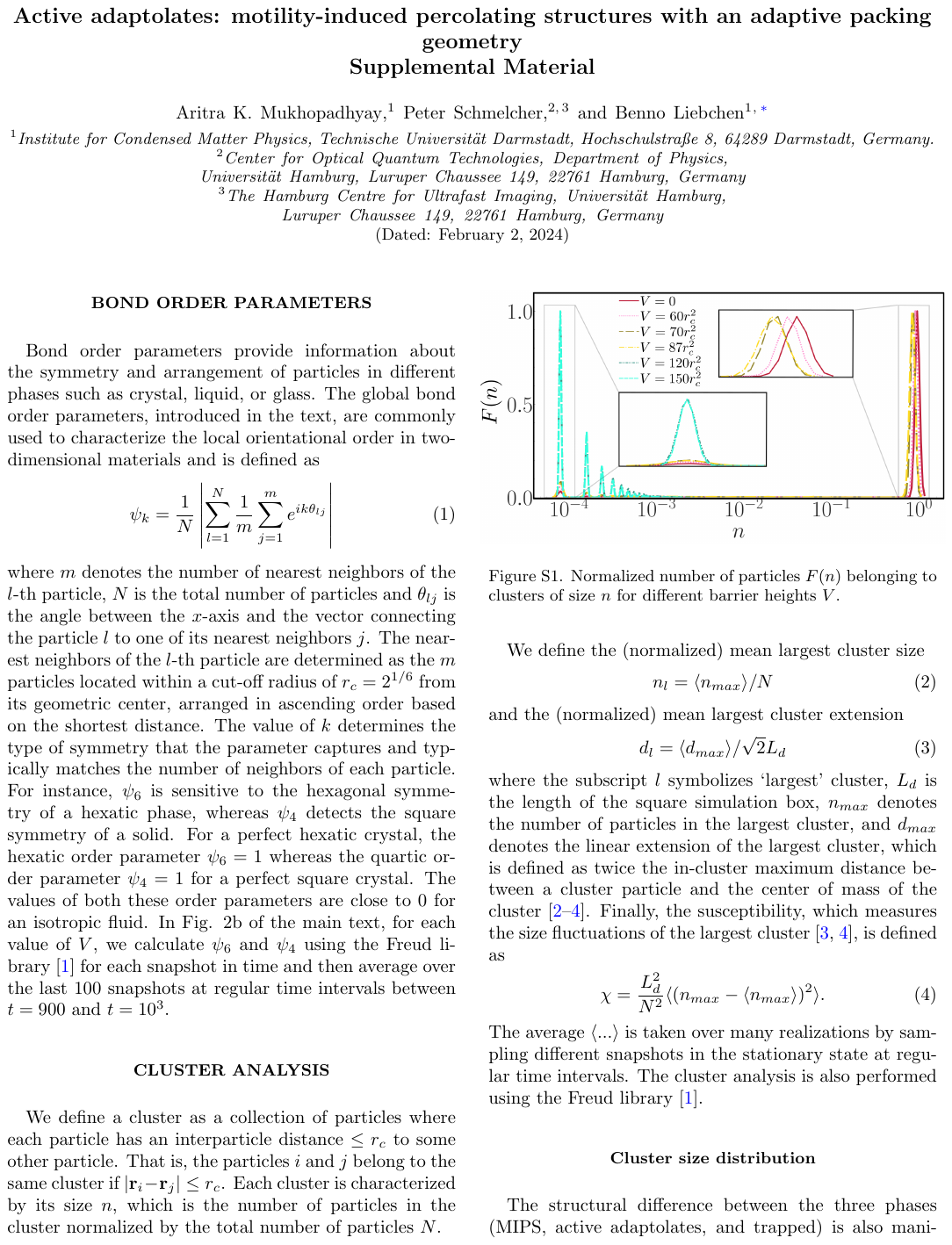}

\pdfximage{\supplementfilename}
\def\numbersupplementpages{\the\pdflastximagepages}

\newif\ifarXiv
\arXivtrue 

\begin{document}
\title{Active adaptolates: motility-induced percolating structures with an adaptive packing geometry}
\author{Aritra K. Mukhopadhyay}
\affiliation{Institute for Condensed Matter Physics, Technische Universit{\"a}t Darmstadt, Hochschulstraße 8, 64289 Darmstadt, Germany.}
\author{Peter Schmelcher}
\affiliation{Center for Optical Quantum Technologies, Department of Physics, Universit\"at Hamburg, Luruper Chaussee 149, 22761 Hamburg, Germany}
\affiliation{The Hamburg Centre for Ultrafast Imaging, Universit\"at Hamburg, Luruper Chaussee 149, 22761 Hamburg, Germany}
\author{Benno Liebchen}
\email{benno.liebchen@pkm.tu-darmstadt.de}
\affiliation{Institute for Condensed Matter Physics, Technische Universit{\"a}t Darmstadt, Hochschulstraße 8, 64289 Darmstadt, Germany.}

\date{\today}

\begin{abstract}
It is well known that periodic potentials can be used to induce freezing and melting in colloids. Here, we transfer this concept to active systems and find the emergence of a so-far unknown active matter phase in between the frozen solid-like phase and the molten phase. This phase of ``active adaptolates" adopts the geometry of the underlying lattice like the frozen phase, maintains ballistic dynamics like the molten phase, and percolates. In particular, this finding creates a route to use external fields for designing the intrinsic structure of active systems without qualitatively affecting their dynamics.
\end{abstract}


\maketitle

\paragraph{Introduction.\textemdash} Phase transitions that are induced by external potentials play an important role in physics. For example, in atomic and condensed matter physics, the superfluid-Mott-insulator transition occurs when changing the depth of a standing light wave (optical lattice) which serves as a periodic potential for (ultracold) atoms \cite{Jaksch_PRL_Cold_Bosonic_1998, Greiner_N_Quantum_Phase_2002,  Bloch_RMP_Manybody_Physics_2008}. This transition separates a superfluid phase, where the atoms are delocalized, occurring in shallow lattices, from an insulating phase, where the atoms are all localized in a steep lattice. Remarkably, this phase transition occurs even at zero temperature and is controlled by the competition between the optical lattice and interaction-controlled quantum fluctuations. In soft matter physics, where thermal fluctuations are important and quantum fluctuations are negligible, the phenomenon of laser-induced freezing and melting uses a similar optical lattice to control a phase transition in colloidal suspensions \cite{Chowdhury_PRL_LaserInduced_Freezing_1985, Das_CS_Laserinduced_Freezing_2001, Wei_PRL_Experimental_Study_1998}. Here, a disordered (gas-like) phase occurs for shallow lattices where colloidal particles diffuse freely, whereas, for steep lattices, we find a solid-like phase in which the particles are confined to the minima of the lattice and adopt the structure of the latter \cite{Reichhardt_PRL_Novel_Colloidal_2002, Brunner_PRL_Phase_Behavior_2002}.

Unlike these equilibrium transitions, substrate- or lattice-induced phase transitions have not yet been explored much in far-from-equilibrium systems. In the present work, we transfer the concept of lattice-induced freezing and melting to active matter systems \cite{Bechinger_RMP_Active_Particles_2016, DeMagistris_PASMaiA_Introduction_Physics_2015, Ramaswamy_ARCMP_Mechanics_Statistics_2010, Zottl_ARCMP_Modeling_Active_2023, Zottl_JPCM_Emergent_Behavior_2016, Liebchen_JPCM_Interactions_Active_2021}, comprising self-propelled particles such as synthetic active colloids \cite{Howse_PRL_SelfMotile_Colloidal_2007, Wang_SM_Practical_Guide_2020, Ginot_NC_Aggregationfragmentation_Individual_2018, Buttinoni_PRL_Dynamical_Clustering_2013, Palacci_S_Living_Crystals_2013}, droplet swimmers \cite{Hokmabad_PRX_Emergence_Bimodal_2021, Izzet_PRX_Tunable_Persistent_2020, Maass_ARCMP_Swimming_Droplets_2016, Michelin_ARFM_SelfPropulsion_Chemically_2023, Feng__Selfsolidifying_Active_2023}, granular microflyers \cite{Scholz_NJP_Ratcheting_Tumbling_2016, Kudrolli_PRL_Swarming_Swirling_2008, Dauchot_PRL_Dynamics_SelfPropelled_2019, Walsh_SM_Noise_Diffusion_2017}, or bacteria \cite{Aranson_RPP_Bacterial_Active_2022, Elgeti_RPP_Physics_Microswimmers_2015, Koch_ARFM_Collective_Hydrodynamics_2011}. For shallow lattices, perhaps unsurprisingly, we find that the active particles undergo motility-induced phase separation (MIPS) \cite{Cates_ARCMP_MotilityInduced_Phase_2015, Digregorio_PRL_Full_Phase_2018, Stenhammar_PRL_Continuum_Theory_2013, Turci_PRL_Phase_Separation_2021, Anderson_NC_Social_Interactions_2022, Mandal_PRL_MotilityInduced_Temperature_2019, Caprini_PRL_Spontaneous_Velocity_2020, Klamser_NC_Thermodynamic_Phases_2018} beyond a critical packing fraction, even when interacting purely repulsively, and self-organize into clusters featuring a liquid-like local hexagonal packing structure. These clusters grow over time and eventually result in a phase-separated state. For steep lattices, similar to passive colloids, we observe a trapping phase where the active particles are trapped in the lattice minima, showing a localized on-site motion. Strikingly, however, for intermediate lattice heights, we find a so-far unknown dynamic percolated state that is induced by activity. In this phase, the particles aggregate locally, similarly as in the early stage of MIPS, but do not show a hexagonal structure. Instead, the aggregates adopt the structure of the underlying square lattice while continuously being in motion, which is reflected by a non-saturating mean-squared displacement. In addition, shortly after emerging, the aggregates merge into a highly dynamic and interconnected percolated structure, distinct from both the trapped and molten phases in terms of large-scale structure. Following their intrinsic active dynamics, local structural adaptation to the substrate lattice, and their percolated large-scale structure, we introduce the portmanteau ``active adaptolates" phase for later convenience. We stress that their adaptivity to the substrate and associated local structures clearly distinguish them from other percolating structures in active matter \cite{Kyriakopoulos_PRE_Clustering_Anisotropic_2019, Sanoria_PRE_Percolation_Transition_2022}. Notably, the phase of active adaptolates is separated from the MIPS and the trapped phases by distinct peaks in the susceptibility, suggesting that it emerges through a proper phase transition. Overall, the present work shows that the idea of laser-induced freezing and melting of colloidal systems can not only be transferred to active matter but also leads to an intriguing intermediate phase, which unifies the dynamic properties of an active liquid with the structural properties of the trapped phase. This can be used in the future to create active fluids with a structure that can be controlled by an external potential.

\begin{figure*}
	\centerline{\includegraphics[width=1.3\textwidth]{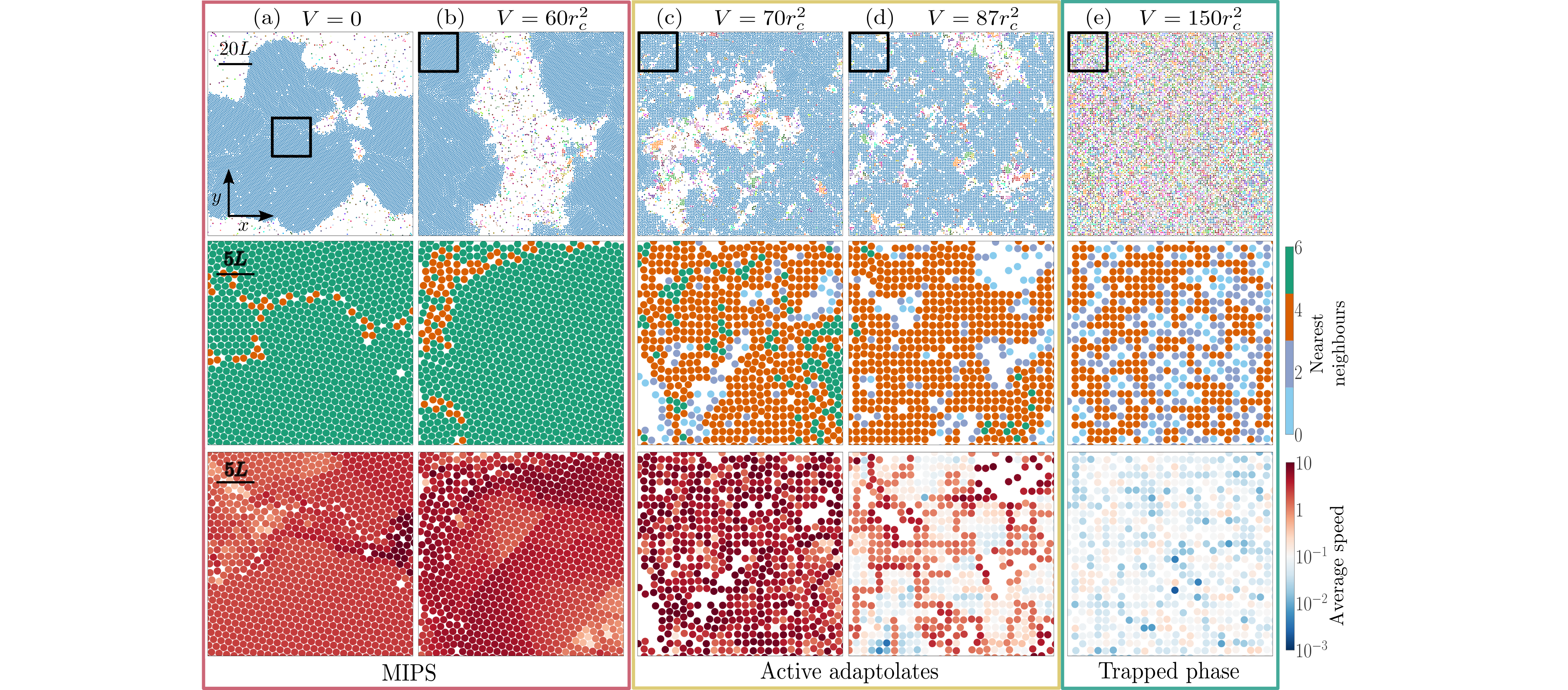}}
	\caption{Lattice-induced freezing and melting in active systems lead to active adaptolates. Upper panel: Steady-state snapshots for (a-e) increasing values of the dimensionless barrier height $V$. Each colored disc denotes an ABP with the color indicating their cluster-ID, i.e. the index of the cluster they belong to. Middle panel: Zoomed snapshots (indicated by the black squares in the upper panel) show the local packing geometry. Lower panel: Zoomed snapshots showing the particle speed, averaged over 100 frames in the steady-state. Parameters: $Pe=300r_c$, $\phi=0.5$, $L=r_c$ and $\epsilon=300r_c^2$.}\label{fig1}
\end{figure*}

\paragraph{Model.\textemdash} We consider a two-dimensional (2D) system of $N$ interacting overdamped active Brownian particles (ABPs). We denote the position and orientation of the $i$-th colloid by $\mathbf{r}'_i$ and $\theta_i$ respectively. Each particle has a diameter $\sigma$, self-propels with velocity $v_0$, and has a translational diffusion coefficient $D_t$. We also fix the rotational diffusion coefficient to $D_r=3D_t/\sigma^2$ satisfying the fluctuation-dissipation theorem in Newtonian equilibrium solvents. The interaction between the particles is modeled by the Weeks-Chandler-Andersen potential characterized by the potential depth $\epsilon'$ and (soft) particle diameter $\sigma$ \cite{Weeks_JCP_Role_Repulsive_1971}. Additionally, we impose an external periodic potential landscape with a spatial periodicity $L'$ in both directions and potential height $V'$. Choosing the units of length and time as $r_u=\sigma$ and $t_u=\sigma^2/D_t$ (three times the persistence time $1/D_r$ of the ABPs) respectively, the dimensionless position $\mathbf{r}_i$ and orientation $\theta_i$ of the $i$-th ABP evolves according to
\begin{eqnarray} \label{abp_eqn}
	\dot {\mathbf{r}}_i&=& Pe\ {\bf p}_i  + {\bf F}_i^{lat} + {\bf F}_i^{int} + \sqrt{2} \pmb{\xi}_i(t) \nonumber \\
	\dot{\theta}_i&=&\sqrt{6} \eta_i (t)
\end{eqnarray}
Here ${\bf p}_i=(\cos \theta_i, \sin \theta_i)$ is the orientation vector of the $i$-th ABP and $Pe=v_0\sigma/D_t$ is the P\'eclet number. The reduced force on the ABPs due to the external lattice is given by ${\bf F}_i^{lat}=-\nabla U (x,y)$ with $U(x,y)=V[\cos ^2 (\pi x/L) + \cos ^2 (\pi y/L)]$; where $V=V'/(k_BT)$ and $L=L'/\sigma$ denote the dimensionless lattice height and spatial periodicity, respectively. $\pmb{\xi}_i(t)$ and $\eta_i (t)$ denote fluctuations modeled by Gaussian white noise with zero mean and unit variance. The reduced interaction force is denoted by ${\bf F}_i^{int}=-\sum_{j\neq i} \nabla U_{ij}$ with $U_{ij}= 4\epsilon \left(\left( 1/r_{ij} \right)^{12} - \left( 1/r_{ij} \right)^{6}\right) + \epsilon$, if $r_{ij}<r_c$ and zero otherwise, where $\epsilon=\epsilon'/(k_BT)$, $r_c=2^{1/6}$ and $r_{ij}=|\mathbf{r}_i - \mathbf{r}_j|$. Experimentally, this model could be realized, for e.g., using micron-sized active colloids that are exposed to a 2D optical lattice \cite{Chowdhury_PRL_LaserInduced_Freezing_1985, Loudiyi_PASMaiA_Direct_Observation_1992, Zemanek_OTOMI_Behavior_Submicron_2005, Buttinoni_E_Active_Colloids_2022} created with a CO$_2$ laser. An alternative realization could be to use granular particles on a 3D-printed periodic substrate mounted on a vibrating plate \cite{Scholz_NJP_Ratcheting_Tumbling_2016, Walsh_SM_Noise_Diffusion_2017}. We numerically integrate \mbox{Eqs.~(\ref{abp_eqn})} using a forward Euler-Maruyama scheme with the LAMMPS open source package \mbox{\cite{Plimpton_JoCP_Fast_Parallel_1995, Thompson_CPC_LAMMPS_Flexible_2022}} for about $1.25\times 10^4$ particles using a timestep $dt=10^{-6}$ for a total time of $t_{tot}=2\times 10^3$ in a square domain of size $L_d \times L_d$ with periodic boundary conditions and $L_d=125r_c$. Our system is characterized by four dimensionless parameters $Pe$, $V$, $L$, $\epsilon$, and the packing fraction of the particles $\phi=N\pi/4L_d^2$. To be in the regime where the system shows MIPS, we fix $Pe=300r_c$ and $\phi=0.5$, and to avoid significant particle overlap, we fix $\epsilon=300r_c^2$. Lastly, to explore the competition between the lattice potential and the inherent tendency of ABPs towards hexagonal packing, we set $L=r_c$ and investigate the collective dynamics of ABPs for different values of $V$.

\begin{figure}
	\centerline{\includegraphics[width=0.66\textwidth]{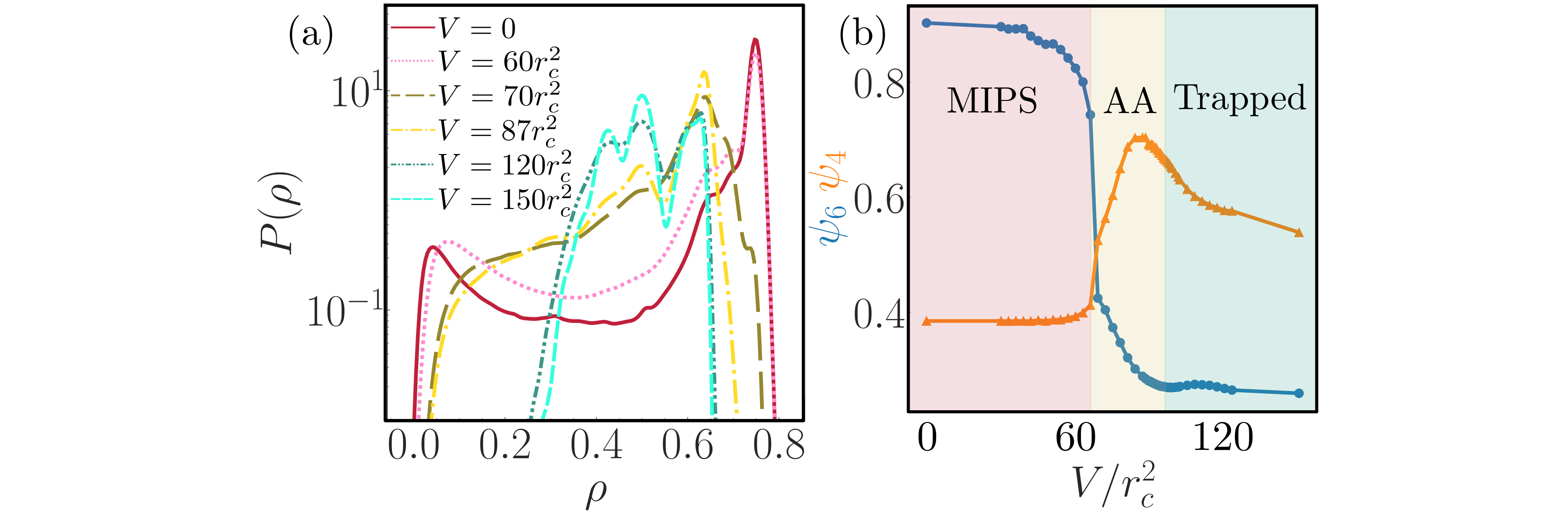}}
	\caption{(a) Distribution $P(\rho)$ of the local density $\rho$ of ABPs for different barrier heights $V$. (b) Global bond order parameters $\psi_4$ (orange triangles) and $\psi_6$ (blue circles) as a function of $V$, averaged over different snapshots in the stationary state. The boundaries between MIPS, active adaptolates (AA), and the trapped phase correspond to the peaks of the susceptibility (Fig.~\ref{fig3}a).}\label{fig2}
\end{figure}

\paragraph{Active adaptolates.\textemdash} In the absence of a periodic potential, the ABPs undergo MIPS and self-organize into hexagonally-packed clusters, defined as collections of particles where each particle has an interparticle distance $\leq r_c$ to some other particle (Fig.~\ref{fig1}a). These clusters grow with time, ultimately coarsening towards a phase-separated state of a coexisting dilute gas-like phase and a dense liquid-like phase \cite{Tailleur_PRL_Statistical_Mechanics_2008, Cates_E_When_Are_2013, Cates_ARCMP_MotilityInduced_Phase_2015, Speck_PRL_Effective_CahnHilliard_2014, Hecht_PRL_Active_Refrigerators_2022, Maloney_SM_Clustering_Phase_2020, Ma_JCP_Dynamical_Clustering_2022, Paoluzzi_CP_Motilityinduced_Phaseseparation_2022, Buttinoni_PRL_Dynamical_Clustering_2013}. As long as the effective force due to the external potential is weaker compared to the effective self-propulsion ($V\pi/L \lesssim 2Pe/3 $, or $V\lesssim 65 r_c^2$), we observe a similar phase-separation of the ABPs into a dilute phase and a dense phase possessing a local hexagonal structure (Fig.~\ref{fig1}b). In contrast, a steep barrier height ($V\pi/L \gtrsim Pe $ or $V\gtrsim 95 r_c^2$) completely suppresses particle aggregation and freezes the particle dynamics similarly as in passive colloids \cite{Chowdhury_PRL_LaserInduced_Freezing_1985, Das_CS_Laserinduced_Freezing_2001}. The ABPs are then localized within potential minima, with each particle occupying a single site, thus yielding a trapped phase similar to passive colloids \cite{Reichhardt_PRL_Novel_Colloidal_2002, Brunner_PRL_Phase_Behavior_2002} (Fig.~\ref{fig1}e). Interestingly, however, at intermediate barrier heights ($65 r_c^2 \lesssim V\lesssim 95 r_c^2$), we observe an additional phase that has no counterpart in classical laser-induced freezing and melting scenarios. In particular, we find that the ABPs rapidly aggregate in large porous network-like structures that span the entire system (Fig.~\ref{fig1}c,d and Movie M1 in Supplemental Material \cite{note1}). Locally, unlike MIPS, they do not show hexagonal packing but adopt the structure of the underlying periodic potential such that the clusters exhibit a dominant square ordering (Figs.~\ref{fig1}d, \ref{fig2}b). Interestingly, these square-ordered percolating structures are still continuously in motion (Movie M2 in Supplemental Material \cite{note1}) and exhibit active diffusion at late times (Fig.~\ref{fig4}). Hence, we call these structures `active adaptolates'.

\paragraph{Structural characterization of different phases.\textemdash} To characterize the observed phases, we examine the distribution of the local density $\rho$ of the ABPs for different lattice heights (Fig.~\ref{fig2}a). We partition the system into Voronoi cells, each cell containing a single ABP, and compute the local density by determining the ratio of the ABP's occupied area to its corresponding Voronoi cell area. In the absence of the lattice and for low lattice height ($V=60 r_c^2$), we observe a bimodal local density distribution, signifying MIPS \cite{Stenhammar_SM_Phase_Behaviour_2014, Klamser_NC_Thermodynamic_Phases_2018}. These peaks at $\rho\sim 0.05$ and $\rho \sim 0.72$ correspond to the co-existing low-density gas-like and high-density liquid-like (hexagonal packing) phases, respectively. The local structure of these phases can be characterized using the global hexatic $\psi_6$ and quartic $\psi_4$ bond order parameters (see Supplemental Material \cite{note1}) whose values are 1 for perfect hexatic and square crystals, respectively \cite{Klamser_NC_Thermodynamic_Phases_2018, Digregorio_PRL_Full_Phase_2018}. For $V\lesssim 60 r_c^2$, $\psi_6>0.8$ (Fig.~\ref{fig2}b) which indicates that the ABPs are hexagonally packed within the dense clusters.

At intermediate barrier heights ($65 r_c^2 \lesssim V\lesssim 95 r_c^2$), the distinct peak at $\rho\sim 0.05$ vanishes (Fig.~\ref{fig2}a), indicating that the active adaptolates are not phase separated. Instead, the density distribution for $V=70 r_c^2$ features a single broad maximum at $\rho \sim 0.62$, which corresponds to the square packing density ($=\pi/4 r_c^2$) of the ABPs in the square lattice. As $V$ is increased further to $V=87 r_c^2$, the peak becomes narrower, indicating that most of the ABPs are now arranged in a square packing structure. This is consistent with the behavior of the bond order parameters in Fig.~\ref{fig2}b, where we see that $\psi_6$ decreases sharply as $V$ is increased beyond $60 r_c^2$ accompanied by a simultaneous increase of $\psi_4$. Overall, we have a lattice-induced transition from a phase-separated state with a hexagonal packing structure to a phase that adopts the structure of the underlying lattice while still requiring the interplay of activity and collision to exist. When $V$ increases further, beyond $120 r_c^2$, the peak height at $\rho\sim 0.62$ decreases (Fig.~\ref{fig2}a) because the square-ordered aggregates are gradually destroyed. Then, the ABPs are randomly and uniformly pinned to different lattice minima because they do not self-propel fast enough to cross the potential barriers. As a result, the local density of the ABPs decreases, and we observe multiple peaks emerging near $\rho\sim \phi = 0.5$ (which is the mean density of the ABPs) corresponding to ABPs with less than four nearest neighbors at the adjacent minima. Since the square packing of the ABPs weakens due to the random filling of the square lattice, $\psi_4$ also decreases (Fig.~\ref{fig2}b). 

\begin{figure} 
	\centerline{\includegraphics[width=0.7\textwidth]{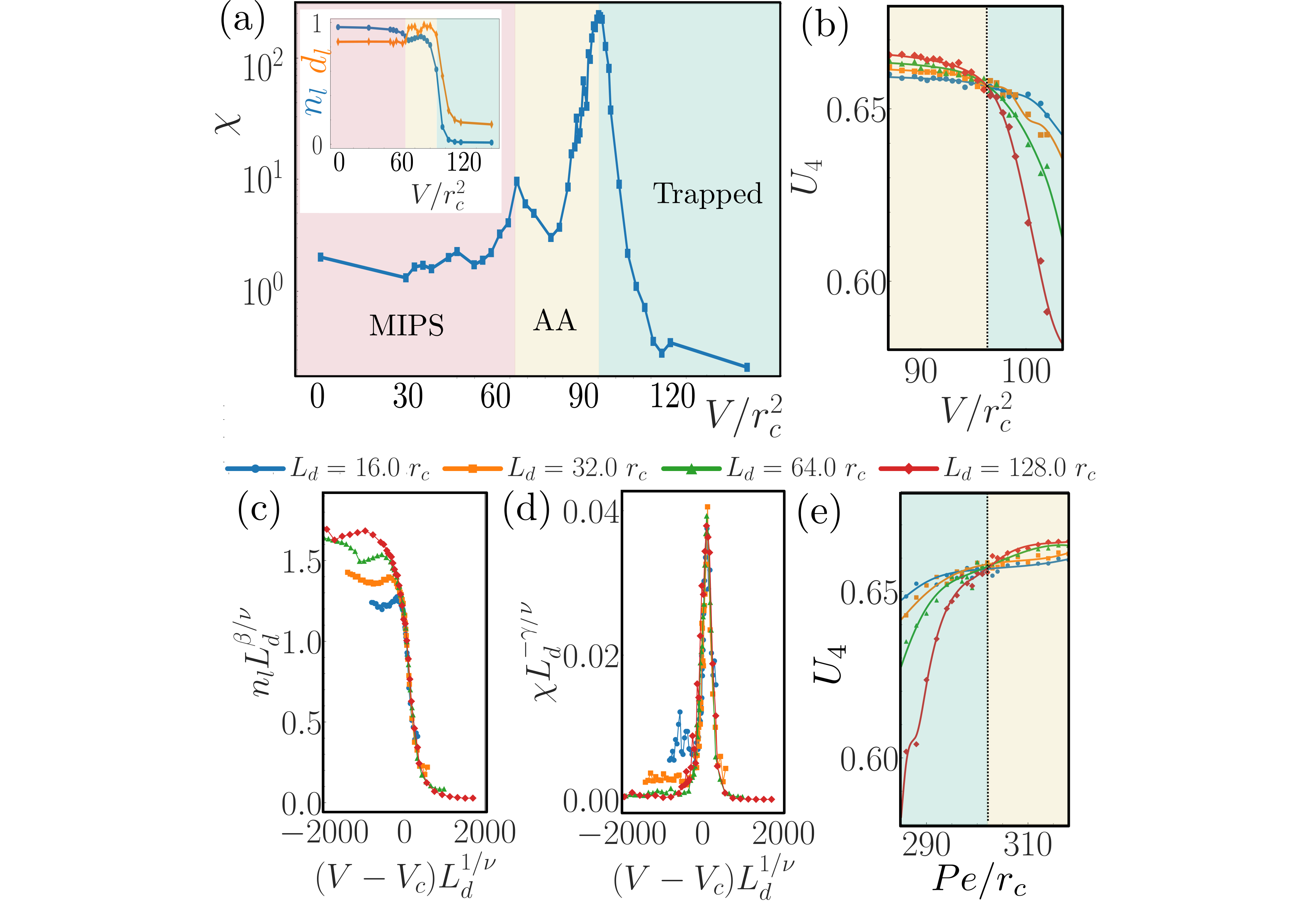}}
	\caption{(a) Susceptibility as a function of the reduced lattice height $V/r_c^2$ showing the three different nonequilibrium phases in the background. The peaks in the susceptibility $\chi$ at $V\approx 65r_c^2$ and $\approx 95r_c^2$ denote the separation of the active adaptolate phase from the MIPS and the trapped phase respectively. Inset: Normalized mean largest cluster size $n_l$ (blue circles) and normalized mean largest cluster extension $d_l$ (orange checks). (b,e) Binder cumulant $U_4$ as a function of $V/r_c^2$ and $Pe/r_c$ respectively for four different system sizes $L_d$. (c,d) Finite-size scaling of $n_l$ and $\chi$ for different $L_d$. The collapse occurs for the critical exponents $\beta\approx 0.16$, $\gamma\approx 2.35$, $\nu\approx 1.25$ and critical lattice height $V_c \approx 96.3r_c^2$, indicating that the transition between the AA and trapped phase is a proper phase transition that belongs to the 2D percolation universality class.} \label{fig3}
\end{figure}

\paragraph{Fluctuations and phase boundaries.\textemdash} To characterize the transition from the MIPS and trapped phase to the active adaptolate phase, we determine the (normalized) mean largest cluster size $n_l$, the (normalized) mean largest cluster extension $d_l$ and the susceptibility $\chi$ measuring the size fluctuations of the largest cluster (see Supplemental Material \cite{note1}) \cite{Levis_PRE_Clustering_Heterogeneous_2014, Kyriakopoulos_PRE_Clustering_Anisotropic_2019, Sanoria_PRE_Percolation_Transition_2022}. $d_l$ quantifies the compactness of the largest cluster; the larger the value of $d_l$, the less compact a cluster is. In the MIPS regime, $n_l\approx 1$ and $d_l\approx 0.8$ (Fig.~\ref{fig3}a, inset) indicate that the ABPs form a large dense cluster. At intermediate lattice heights, where active adaptolates are formed, $n_l$ slightly decreases to $\approx 0.8$ whereas $d_l$ increases to its maximum value $\approx 1$. This shows that the MIPS phase undergoes a qualitative structural change at intermediate lattice heights to form a less compact porous percolated space-filling cluster in which the local packing of the particles adopts the square geometry of the underlying lattice. For $V\gtrsim 95r_c^2$, the percolated cluster breaks up into many smaller clusters (see Fig.~S1 in Supplemental Material \cite{note1}), leading to a drastic decrease in the values of $n_l$ and $d_l$. The distinct peaks in susceptibility $\chi$ at $V\approx 65r_c^2$ and $\approx 95r_c^2$ indicate that the active adaptolate phase is separated from both the MIPS and the trapped phase by a phase transition, rather than emerging gradually in the form of a crossover. 

To verify this, we perform a finite-size scaling analysis for different system sizes $L_d=16r_c,32r_c,64r_c$, and $128r_c$ with $V$ as the control parameter. For a percolation phase transition, the order parameter $n_l$ and susceptibility $\chi$ are expected to scale as $n_l(L_d)=L_d^{-\beta/\nu}f\left(|V-V_c|L_d^{1/\nu} \right)$ and $\chi(L_d)=L_d^{\gamma/\nu}g\left(|V-V_c|L_d^{1/\nu} \right)$ for different $L_d$ \mbox{\cite{Kyriakopoulos_PRE_Clustering_Anisotropic_2019,Gawlinski_JPAMG_Continuum_Percolation_1981,Heermann_ZPB-CM_Influence_Boundary_1980}}. Here $V_c$ is the critical lattice potential height at the transition point, $f$ and $g$ are the scaling functions, and $\beta$, $\gamma$, and $\nu$ are the critical exponents. We calculate the fourth-order Binder cumulant $U_4=1-\langle d_l^4\rangle/3\langle d_l^2\rangle^2$ of the order-parameter $d_l$ for different $L_d$ \mbox{(Fig.~\ref{fig3}b)}. The crossing point of these curves gives the critical parameter $V_c$ \mbox{\cite{Binder_PRL_Critical_Properties_1981,Binder__Monte_Carlo_2019}}, which we estimate to be $\approx 96.3r_c^2$. We estimate the critical exponents to be $\beta\approx 0.16$, $\gamma\approx 2.35$, and $\nu\approx 1.25$ (see Supplemental Material \mbox{\cite{note1}}) which are in close agreement with the known universal exponents for the 2D percolation transition ($\beta\approx 0.14$, $\gamma\approx 2.38$, $\nu\approx 1.33$) \mbox{\cite{Gawlinski_JPAMG_Continuum_Percolation_1981}}. Using these values and $V_c\approx 96.3r_c^2$, we find a collapse of $n_l$ and $\chi$ \mbox{(Fig.~\ref{fig3}c,d)} at the critical point $V_c$ for different system sizes $L_d$, suggesting that the transition from the adaptolate to the trapped phase is indeed a phase transition that belongs to the (static) percolation universality class. In addition, we stress that the adaptolate phase features significant, ongoing intrinsic dynamics (active diffusion at late times); see \mbox{Fig.~\ref{fig4}}. 

To show that the active adaptolate phase is activity-induced, we also performed a finite-size scaling analysis with the P\'eclet number $Pe$ as the control parameter (and fixing $V=V_c$) and observed that active adaptolates only emerge when the activity surpasses a critical threshold, $Pe_c\approx 302.3r_c$ \mbox{(Fig.~\ref{fig3}e)}, showing that activity is crucial to realize such percolating dynamic clusters that adopt the structure of the underlying potential landscape (see \mbox{Fig.~S2} in Supplemental Material \mbox{\cite{note1}} for finite size scaling of $n_l$ and $\chi$ for different $Pe$).

\begin{figure} 
	\centerline{\includegraphics[width=0.63\textwidth]{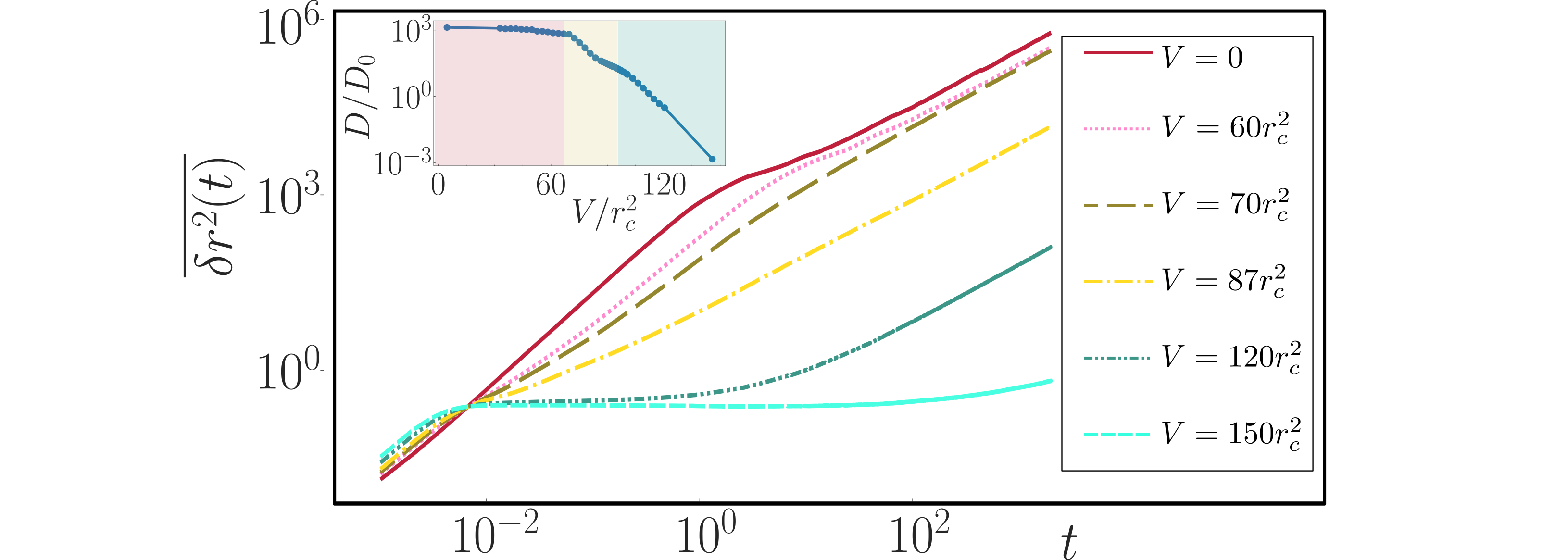}}
	\caption{Mean squared displacement of the ABPs as a function of time $t$ for different $V$. Inset: Late-time diffusion coefficient $D$ of the ABPs, normalized by the diffusion coefficient $D_0$ of passive particles for $V=0$, as a function of $V$. Notice that $D$ decays much slower and smoother than $\psi_6$ (Fig.~\ref{fig2}b). The figure indicates that the particles in the active adaptolate phase move one to two orders of magnitude faster than free passive Brownian particles.} \label{fig4}
\end{figure}

\paragraph{Dynamical characterization of different phases.\textemdash} We characterize the dynamical properties of the different phases based on the mean squared displacement (MSD) $\overline{\delta r^2 (t)}  = \frac{1}{N}\sum_{i=1}^{N} \left(\mathbf{r}_i(t)-\mathbf{r}_i(0)\right)^2$ of the ABPs. For $V\lesssim 65 r_c^2$ (MIPS phase), the ABPs exhibit a super-diffusive behavior at intermediate times with $\overline{\delta r^2 (t)} \sim t^2$, followed by a diffusive behavior with $\overline{\delta r^2 (t)} \sim t$ at longer times (Fig.~\ref{fig4}). Even at intermediate barrier heights of $V=70 r_c^2$ and $V=87 r_c^2$, where square-ordered active adaptolates are formed, the ABPs move diffusively with $\overline{\delta r^2 (t)} \sim t$ at late times with a late-time diffusion coefficient $D=\lim\limits_{t \to \infty} \frac{1}{4}\frac{d\ \overline{\delta r^2 (t)}}{dt}$ \cite{Zeitz_EPJE_Active_Brownian_2017} (Fig.~\ref{fig4}, inset) that is about $10-10^2$ times larger than for passive particles. This shows that although the percolating aggregates adopt the underlying structure of the lattice, they are highly dynamic and exhibit dynamical properties similar to active particles in the absence of the lattice. A steeper barrier height decreases the long-time MSD of the ABPs and for $V=150 r_c^2$ their dynamics is almost frozen on the timescale of our simulations, leading to the trapped phase.


\paragraph{Conclusions.\textemdash} This work shows that the concept of laser-induced freezing and melting is not limited to passive systems, but can be extended to active colloids, where it results in a so-far unknown phase of active matter. This phase requires activity to emerge and is separated from the trapped and molten phases by distinct peaks in the susceptibility, i.e. by sharp transitions. As its defining feature, this phase is characterized by a system-spanning percolated large-scale structure with a local packing geometry that adopts the structure of an underlying periodic potential and persistent active diffusive dynamics. Overall, the present work opens the route toward the creation of active matter states whose local intrinsic structure can be controlled with external fields without suppressing their characteristic dynamics. This could potentially be used in the future to steer the optical and mechanical properties of active materials with standing light waves or other spatially periodic or aperiodic potentials.


\begin{acknowledgments}
We acknowledge financial support from the Deutsche Forschungsgemeinschaft (DFG, German Research Foundation) through project number 233630050 (TRR-146).
\end{acknowledgments}

\bibliography{mybib}

\ifarXiv


\section*{Supplementary material}

# Active adaptolates: motility-induced percolating structures with an adaptive packing geometry
# Supplemental Material

Aritra K. Mukhopadhyay,[1] Peter Schmelcher,[2,3] and Benno Liebchen[1,*]

[1]*Institute for Condensed Matter Physics, Technische Universität Darmstadt, Hochschulstraße 8, 64289 Darmstadt, Germany.*
[2]*Center for Optical Quantum Technologies, Department of Physics,
Universität Hamburg, Luruper Chaussee 149, 22761 Hamburg, Germany*
[3]*The Hamburg Centre for Ultrafast Imaging, Universität Hamburg,
Luruper Chaussee 149, 22761 Hamburg, Germany*

(Dated: February 2, 2024)

## BOND ORDER PARAMETERS

Bond order parameters provide information about the symmetry and arrangement of particles in different phases such as crystal, liquid, or glass. The global bond order parameters, introduced in the text, are commonly used to characterize the local orientational order in two-dimensional materials and is defined as

$$\psi_k = \frac{1}{N} \left| \sum_{l=1}^{N} \frac{1}{m} \sum_{j=1}^{m} e^{ik\theta_{lj}} \right| \quad (1)$$

where $m$ denotes the number of nearest neighbors of the $l$-th particle, $N$ is the total number of particles and $\theta_{lj}$ is the angle between the $x$-axis and the vector connecting the particle $l$ to one of its nearest neighbors $j$. The nearest neighbors of the $l$-th particle are determined as the $m$ particles located within a cut-off radius of $r_c = 2^{1/6}$ from its geometric center, arranged in ascending order based on the shortest distance. The value of $k$ determines the type of symmetry that the parameter captures and typically matches the number of neighbors of each particle. For instance, $\psi_6$ is sensitive to the hexagonal symmetry of a hexatic phase, whereas $\psi_4$ detects the square symmetry of a solid. For a perfect hexatic crystal, the hexatic order parameter $\psi_6 = 1$ whereas the quartic order parameter $\psi_4 = 1$ for a perfect square crystal. The values of both these order parameters are close to 0 for an isotropic fluid. In Fig. 2b of the main text, for each value of $V$, we calculate $\psi_6$ and $\psi_4$ using the Freud library [1] for each snapshot in time and then average over the last 100 snapshots at regular time intervals between $t = 900$ and $t = 10^3$.

## CLUSTER ANALYSIS

We define a cluster as a collection of particles where each particle has an interparticle distance $\leq r_c$ to some other particle. That is, the particles $i$ and $j$ belong to the same cluster if $|\mathbf{r}_i - \mathbf{r}_j| \leq r_c$. Each cluster is characterized by its size $n$, which is the number of particles in the cluster normalized by the total number of particles $N$.

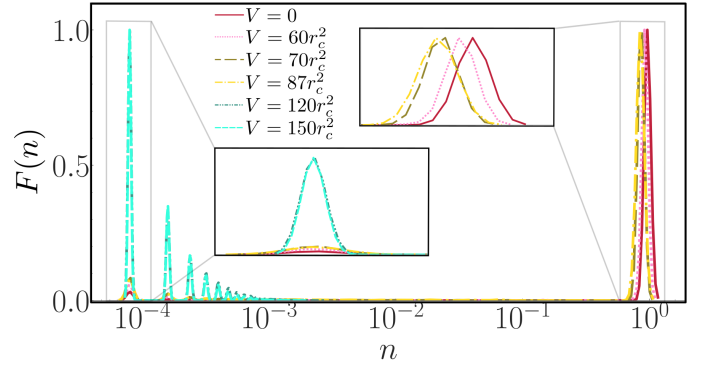

Figure S1. Normalized number of particles $F(n)$ belonging to clusters of size $n$ for different barrier heights $V$.

We define the (normalized) mean largest cluster size

$$n_l = \langle n_{max} \rangle / N \quad (2)$$

and the (normalized) mean largest cluster extension

$$d_l = \langle d_{max} \rangle / \sqrt{2} L_d \quad (3)$$

where the subscript $l$ symbolizes 'largest' cluster, $L_d$ is the length of the square simulation box, $n_{max}$ denotes the number of particles in the largest cluster, and $d_{max}$ denotes the linear extension of the largest cluster, which is defined as twice the in-cluster maximum distance between a cluster particle and the center of mass of the cluster [2–4]. Finally, the susceptibility, which measures the size fluctuations of the largest cluster [3, 4], is defined as

$$\chi = \frac{L_d^2}{N^2} \langle (n_{max} - \langle n_{max} \rangle)^2 \rangle. \quad (4)$$

The average $\langle ... \rangle$ is taken over many realizations by sampling different snapshots in the stationary state at regular time intervals. The cluster analysis is also performed using the Freud library [1].

### Cluster size distribution

The structural difference between the three phases (MIPS, active adaptolates, and trapped) is also mani-

fested in the cluster size distribution [5–7] for different $V$. In Fig. S1, we show the number of particles $F(n)$ belonging to clusters of size $n$ in the stationary state for different values of the lattice height $V$. For better visibility, $F(n)$ is normalized by its maximum value for each $V$. For $V \leq 87 r_c^2$, we see that most of the particles belong to a large cluster of size $n \gtrsim 0.7$. In contrast, only small clusters are formed for steep lattice heights $V \geq 120 r_c^2$. Therefore, $F(n)$ clearly distinguishes the trapped phase observed for steep lattice heights from the MIPS and active adaptolate phases at low and intermediate lattice heights, respectively.

## FINITE-SIZE SCALING

We perform a finite-size scaling analysis for different system sizes $L_d = 16 r_c, 32 r_c, 64 r_c$, and $128 r_c$ with $V$ as the control parameter. For a percolation phase transition, it is known that the order parameter $n_l$ and susceptibility $\chi$ are expected to scale as $n_l(L_d) = L_d^{-\beta/\nu} f\left(|V - V_c| L_d^{1/\nu}\right)$ and $\chi(L_d) = L_d^{\gamma/\nu} g\left(|V - V_c| L_d^{1/\nu}\right)$ for different $L_d$. Here $V_c$ is the critical potential height (barrier) at the transition point, $f$ and $g$ are the scaling functions, and $\beta, \gamma$, and $\nu$ are the critical exponents. The fourth-order Binder cumulant $U_4 = 1 - \langle d_l^4 \rangle / 3 \langle d_l^2 \rangle^2$ of the order-parameter $d_l$ takes the value $\approx 2/3$ in the 'ordered' phase ($d_l \approx 1$, i.e. the active adaptolate phase) and decays to zero in the 'disordered' phase ($d_l \approx 0$, i.e. the trapped phase). We first calculate $V_c$ from the crossing point of the Binder cumulants as shown in Fig. 3b of the main text. Since the derivative of the Binder cumulant scales as $\partial U_4 / \partial V \propto L_d^{1/\nu}$ at the critical point $V = V_c$, we estimate the critical exponent $\nu$ by performing a linear fit of $\log(\partial U_4 / \partial V)$ over $\log L_d$ at $V = V_c$ [8, 9]. Similarly, since the scaling functions $f$ and $g$ are both constants that are independent of the system size at the critical point $V = V_c$, we have $n_l \propto L_d^{-\beta/\nu}$ and $\chi \propto L_d^{\gamma/\nu}$ at $V = V_c$. By performing linear fits of $\log n_l$ and $\log \chi$ with respect to $\log L_d$, we estimate the ratios $\beta/\nu$ and $\gamma/\nu$ respectively, from which we extract the values of the exponents $\beta$ and $\gamma$.

We have also performed a finite-size scaling analysis for different values of activity $Pe$ and a fixed barrier height $V = V_c$. The critical value of the activity was calculated as $Pe_c \approx 302.3 r_c$ (Fig. S2a) and the critical exponents were determined to be $\beta \approx 0.10$, $\gamma \approx 1.73$, $\nu \approx 0.89$, which lead to a good collapse of the order parameter $n_l$ and susceptibility $\chi$ (Figs. S2b,c). This shows that the active adaptolate phase is activity-induced, i.e. it occurs only when the active particles self-propel with a velocity higher than a critical self-propulsion velocity.

We have also explored the robustness of the adaptolate phase for different packing fractions $\phi$. For higher

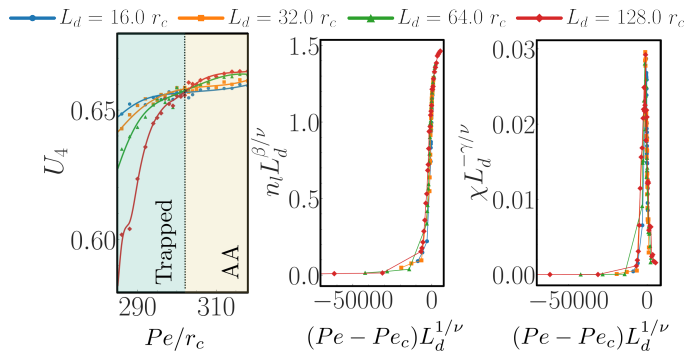

Figure S2. (a) Binder cumulant $U_4$ as a function of $Pe/r_c$ for four different system sizes $L_d$. (b,c) Finite-size scaling of $n_l$ and $\chi$ respectively for different $L_d$.

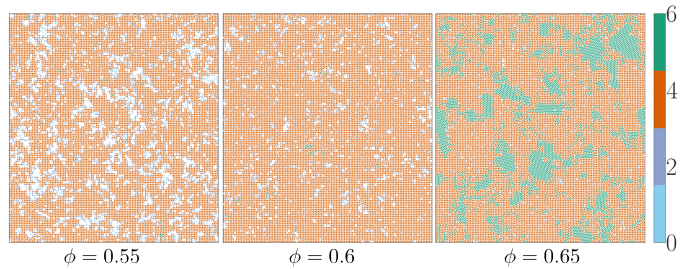

Figure S3. Snapshots of ABPs (active Brownian particles) for different area fractions. The colors denote the number of nearest neighbors. Parameters: $Pe = 104 r_c, V = 96.3 r_c^2$, other parameters same as in Fig.1 in the manuscript.

values of $\phi = 0.55, 0.6$, we observe similar square-ordered dynamic percolating clusters (Fig. S3) as observed for $\phi = 0.5$ in the main text. At even higher packing fraction ($\phi = 0.65$), we still observe dynamic percolating structures. At such relatively high packing fractions, they coexist with isolated domains of hexatically ordered ABPs.

## LIST OF MOVIES

- Movie M1 shows the clustering of the particles in the three different phases, i.e., MIPS, active adptolates, and the trapped phase. The particles are colored by the cluster-ID, which is the index of the cluster they belong to.

- Movie M2 shows the local packing geometry of the particles in the three phases. The colors denote the number of nearest neighbors of each particle.

---


\fi

\end{document}